\documentclass[twocolumn,tightenlines,floats,floatfix,prc,nofootinbib,preprintnumbers]{revtex4}
\def\bea{\begin{eqnarray}}
\def\eea{\end{eqnarray}}

\def\st#1{{\kern-4pt} \not\!#1}

\def\sp{\kern +3pt}
\def\sm{\kern -3pt}

\usepackage[usenames]{color}

\def\be{\begin{equation}}
\def\ee{\end{equation}}
\def\ba{\begin{eqnarray}}
\def\ea{\end{eqnarray}}

\usepackage{graphics}
\usepackage{graphicx}
\usepackage{epsf} 
\usepackage{amsmath}
\usepackage{amssymb}
\usepackage{slashed}
\usepackage{bm}

\def\sfrac#1#2{{\textstyle \frac{#1}{#2}}}

\begin{document}

\phantom{0}
\vspace{-0.2in}
\hspace{5.5in}

%\parbox{1.5in}{ } %\leftline{JLAB-THY-10-1219}}

\preprint{{\bf LFTC-17-9/9 }}

\vspace{-1in}%\parbox{1.5in}{ \vspace{-9.6in}}  % moves the preprint box down

\title
{\bf  Analytic parametrizations of the 
$\gamma^\ast N \to N(1440)$ form factors 
inspired by light-front holography}

\author{G.~Ramalho$^{1,2}$ 
\vspace{-0.1in} }

\affiliation{$^1$Laborat\'orio de F\'{i}sica Te\'orica e Computacional -- LFTC,
Universidade Cruzeiro do Sul, 01506-000, S\~ao Paulo, SP, Brazil
\vspace{-0.15in}}

\affiliation{$^2$International Institute of Physics,
Federal University of Rio Grande do Norte,
Campus Universit\'ario - Lagoa Nova,  CP.~1613
Natal Rio Grande do Norte 59078-970, Brazil}

\vspace{0.2in}
\date{\today}

\phantom{0}

\begin{abstract}
We present analytic parametrizations for the 
$\gamma^\ast N \to N(1440)$ form factors 
derived from light-front holography 
in leading twist approximation.
The new parametrizations describe 
the electromagnetic form factors 
using analytic functions dependent on 
the masses of the $\rho$ meson and 
respective radial excitations,
as well as the masses of the nucleon and the resonance $N(1440)$.  
The free parameters of the model,  
associated with three independent couplings, 
are interpreted as bare couplings, 
and are fixed by the nucleon data for large $Q^2$.
The proposed parametrizations compare remarkably 
well with the empirical data for $Q^2 > 2$ GeV$^2$,
corroborating the dominant role  
of the valence quark degrees of freedom in the 
$\gamma^\ast N \to N(1440)$ transition.
\end{abstract}

%\phantom{0}
%\vspace{7.0in}
%\vspace{-6in}
\vspace*{0.9in}  % sets how far the title is below the preprint box
\maketitle

\section{Introduction}

It was found recently
that the combination of the 5D gravitational 
anti-de-Sitter (AdS) space and conformal field theories (CFT) 
can be used to study strong coupling theories like QCD in the confining 
regime~\cite{Maldacena99,Witten98,Gubser98,Brodsky15}.
This application is possible due to the correspondence 
between the results from AdS/CFT and the 
results from light-front dynamics 
based on a Hamiltonian that includes the confining mechanism of 
QCD~\cite{Brodsky15,Brodsky08a,Teramond09a}.
The correspondence  between AdS/QCD and light-front dynamics 
is the consequence of the mapping description 
of the hadronic modes in AdS space 
and the Hamiltonian formulation of 
QCD in physical space-time quantized on the light-front.
This connection (duality) is referred as Light-Front holography 
or Holographic QCD. 
As a consequence of this duality,  for hadrons with massless quarks,
one can relate the AdS holographic variable $z$
with the impact separation variable $\zeta$,
which measures the distance of the constituent partons 
inside the hadrons,
and calculate wave functions 
of the hadrons in terms of these variables~\cite{Brodsky15,Brodsky08a,Teramond09a}.

Over the past few years 
light-front holography has been used
to study the properties of the hadrons,
such as the mass spectrum,
parton distribution 
functions~\cite{Brodsky15,Karch06a,Grigoryan07a,Chakrabarti14a,Liu15a}, 
and structure form factors of 
mesons and baryons~\cite{Brodsky08a,Abidin09,Gutsche12a,Chakrabarti13a,Teramond11a,Teramond12a,Gutsche13a,paper1,Sufian17a}.

The reduction from the 5D action to 4D 
introduces the holographic mass scale, $\kappa$,
which establishes the scale of the meson 
and baryon spectrum. 
Light-front holography provides a good 
description of the physical spectrum  
when we consider the soft-wall model 
associated with a confinement potential 
$U(z) = \kappa^2 z^2$~\cite{Brodsky15,Teramond12a,Branz10a,Teramond15a}.
The scale $z \sim 1/\kappa$ separates the ultraviolet conformal limit 
$z \to 0$ (perturbative region) 
from the infrared region (large $z$) regulated by the confinement 
(non-perturbative region).
In the limit $\kappa \to 0$ the chiral symmetry is exact; 
when $\kappa$ is very large the conformal symmetry
is broken~\cite{Brodsky15,Teramond12a}.

The study of the electromagnetic structure of 
the baryons can be performed in light-front holography 
noting that the three-quark systems
can be regarded as two-body systems with 
an active quark and a spectator quark-pair, 
with masses and wave functions determined by
the light-front wave equations~\cite{Brodsky15}.
In previous works it was found that light-front holography
provides good estimates 
for the electromagnetic form factors of the pion, 
the nucleon, and the Roper,
when one uses wave functions based on the first  
Fock states~\cite{Gutsche12a,Teramond12a,Gutsche13a,Brodsky15,Sufian17a}.
The inclusion of higher order Fock states
means that corrections to the three-valence quark approximation 
(leading twist approximation), such as the meson cloud effects,
are taken into account. 
It was however found, that good estimates 
for the form factors can also be obtained 
in leading twist approximation~\cite{Abidin09,Chakrabarti13a,Teramond12a,Brodsky15,Brodsky08a,paper1,Teramond11a}.
Although the leading twist approximation may look as 
a rough simulation of the real world,
it may still provide an excellent first estimate,
since the confinement is effectively taken 
into account in the light-front wave functions~\cite{Teramond12a,Brodsky15,Brodsky08a,Liu15a}.

The transition current associated 
with the $\gamma^\ast N \to N^\ast$ transitions, 
where $N^\ast$  is a $J^P= \sfrac{1}{2}^+$ resonance 
(spin 1/2 and positive parity), is  characterized by 
two independent functions of $Q^2$: 
the Dirac (spin-nonflip)
and the Pauli (spin-flip) form factors~\cite{Brodsky80,Brodsky01}.

Brodsky and Teramond have shown that,
in leading twist approximation,
an accurate description of the form factors for the nucleon 
and the $\gamma^\ast N \to N(1440)$ Dirac form factor can be obtained 
using analytic expressions based on 
the $\rho$ meson masses~\cite{Brodsky15,Teramond11a,Teramond12a}.

In the present work, 
we propose simple analytic parametrizations 
for both $\gamma^\ast N \to N(1440)$ transition form factors.
In those parametrizations, 
the $\gamma^\ast N \to N(1440)$ transition form factors are 
expressed in terms of the $\rho$ meson masses, as well as 
the nucleon and Roper masses.
The parametrizations compare well with empirical data
for $Q^2 > 2$ GeV$^2$.
Here $Q^2$ is defined by $Q^2= -q^2$, 
where $q$ is the momentum transfer.

\section{Light-Front holography}

In light-front holography, the electromagnetic 
interaction with the hadrons is defined 
in terms of a 5D action which includes 
the coupling with the electromagnetic field 
$V_M$ ($M=0,1,2,3,z$), and it is at the end reduced to the 4D current 
$J_\mu$~\cite{paper1,Grigoryan07a,Gutsche12a,Abidin09}.
The simplest possible coupling is the minimal Dirac coupling, 
which has the form  $\hat Q \, \Gamma^M V_M$, 
where $\hat Q$ is the charge transition operator and  
$\Gamma^M$ is a 5D gamma matrix. %($M=0,1,2,3,z$).
With the minimal Dirac coupling one obtains 
only contributions for the Dirac form factor~\cite{Abidin09}.
One can generate contributions for the  Pauli form factor when we 
include a non-minimal coupling 
of the form 
$\sfrac{i}{4} \eta_N \left[\Gamma^M, \Gamma^N \right] V_{MN}$
where $\eta_N$ is a parameter associated 
with the the proton ($N=p$) 
or neutron ($N=n$) anomalous magnetic moment, and 
$V_{MN} = \partial_M V_N - \partial_N V_M + [V_M,V_N]$~\cite{Abidin09,Gutsche12a,Brodsky15,Chakrabarti13a,paper1}.
This non-minimal coupling gives also an 
extra contribution for the Dirac form factor~\cite{Abidin09,Chakrabarti13a}.
Other couplings can be introduced in the 5D action.
A new minimal-type coupling was included 
in Refs.~\cite{Gutsche12a,Gutsche13a} 
with the form $g_V  \tau_3 \Gamma^M \gamma_5 V_M$, 
where $g_V$ is an isovector coupling constant 
and $\tau_3$ is the Pauli isospin operator.
This new coupling is absent in the 4D action, 
but can appear in the 5D action, inducing additional contributions 
to the Dirac form factor.
Alternative non-minimal couplings are discussed in Ref.~\cite{newpaper}.

In holography, the calculation of the transition currents 
between two baryon states 
is performed considering the overlap of the 
holographic electromagnetic field, 
which includes the electromagnetic couplings mentioned above, 
with the baryon fields $\Psi_B(x,z)$
associated with the initial and final states.
In this notation  $x$ is the 4D space-time coordinate
and $z$ the holographic variable.
The fields $\Psi_B(x,z)$ are determined by 
equations of motion  that can be reduced 
to Sch\"{o}dinger-type wave equations in the variable $z$.
Check Refs.~\cite{Brodsky15,Abidin09,Gutsche12a,Gutsche13a} for a review.
From the transition currents we can extract 
the holographic expressions for the electromagnetic form factors
which depend on the minimal-type 
coupling ($g_V$) and the two non-minimal couplings 
($\eta_p$ and $\eta_n$), which characterize 
the electromagnetic interaction with the quarks inside the hadrons.
The parameters $g_V$, $\eta_p$ and $\eta_n$ are related with the intrinsic properties 
of the valence quarks and represent therefore bare couplings.

Calculations of the nucleon 
electromagnetic form factors based on 
the Dirac minimal coupling and the non-minimal couplings 
can be found in Refs.~\cite{Brodsky15,Abidin09,Gutsche12a,Chakrabarti13a,Teramond11a,Teramond12a,Sufian17a}.
Calculations that include the 
minimal-type coupling $g_V$ for the nucleon 
and the Roper are presented in Refs.~\cite{Gutsche12a,Gutsche13a, paper1}.

In light-front holography 
the expressions for the baryon electromagnetic form factors 
include functions with poles on $Q^2$
associated with the masses $M^2= 4 (n+1) \kappa^2$, 
where $n=0,1,...$~\cite{Gutsche12a,Gutsche13a,Abidin09,Chakrabarti13a}.
Some authors prefer to represent those functions 
using a pole structure based on 
the vector meson dominance~\cite{Grigoryan07a,Teramond11a,Teramond12a,Brodsky15},
which is expected to appear in high energy processes.
This representation is possible 
because the high energy physical photons 
have hadronic components that can be 
expressed as intermediate particles 
with the quantum numbers of the 
physical vector mesons~\cite{Maldacena99,Sakurai60,Bauer79,Gari84,Lomon01}.

The simultaneous description of the 
vector meson and nucleon radial excitation
spectrum based on only a mass scale $\kappa$
is a difficult task~\cite{Brodsky15}.
The vector meson spectrum is well explained with $\kappa \simeq 0.385$ GeV,
however, the nucleon and Roper masses are only 
roughly approximated by the holographic estimates
in leading twist approximation~\cite{Gutsche12a,Gutsche13a}.
For that reason different representations 
of the baryon transition form factors have been used,
as mentioned above.
One representation keeps the pole structure $M^2= 4 (n+1) \kappa^2$,
and  fixes $\kappa$ by the nucleon mass or the 
$\rho$ mass~\cite{Gutsche12a,Gutsche13a,Abidin09,Chakrabarti14a}. 
In a second representation there is a 
re-interpretation of the poles, 
which are shifted to the vector meson physical poles, 
$M  \to 2 \sqrt{2n+ 1} \kappa$ ($n=0,1,...$),
as proposed by Brodsky and Teramond~\cite{Brodsky15,Teramond12a}.

In the present work we propose an alternative 
representation of the $\gamma^\ast N \to N(1440)$ transition form factors, 
where the mass poles are re-interpreted 
as vector meson poles or as mass poles of the nucleon and Roper.
This re-interpretation is based on the holographic estimates 
of the masses in leading twist approximation, expressed in terms of the mass scale $\kappa$, 
and in the empirical masses of the  nucleon and the nucleon radial excitations.
The proposed expressions are not strictly derived from light-front holography,
but they are still inspired by  light-front holography.

We consider then a bottom-up approximation to QCD,
starting with a well defined interaction Lagrangian, 
and use the phenomenology to correct the expressions
for the mass poles derived from first principles.
Examples of {\it ad hoc} corrections to the light-front holography estimates 
are common in the literature.
The minimal-type coupling ($g_V$) 
and the Pauli coupling ($\eta_N$) 
were introduced for phenomenological purposes~\cite{paper1,Abidin09,Chakrabarti13a}.
The calculation of the nucleon and Roper 
form factors in the model from Brodsky-Teramond~\cite{Teramond11a,Brodsky15}
is improved when we use holographic vector masses 
instead of the empirical masses.
Those corrections to the light-front holography results motivate the following 
representation of the $\gamma^\ast N \to N(1440)$ transition form factors.

\section{$\gamma^\ast N \to N(1440)$ transition form factors}

The $\gamma^\ast N \to N(1440)$ transition form factors
have been calculated using light-front holography
based on the nucleon and the Roper wave functions 
with Fock states up to the twist order $\tau=5$~\cite{Gutsche12a,Gutsche13a}.
The leading twist component ($\tau=3$)
corresponds to the three-quark (3$q$) state; 
the following twist component ($\tau=4$) 
represents the gluon excitation of the three-quark state,  $(3q)g$; 
and the $\tau=5$ correspond to the quark-antiquark excitation  ($q \bar q$) 
of the  three-quark state.
Thus, $\tau=3$ represents the valence quark approximation
and $\tau=5$ takes into account the meson cloud excitation of the three-quark core.

In the leading twist approximation ($\tau=3$) 
one can express the 
$\gamma^\ast N \to N(1440)$ transition form factors in 
the form~\cite{paper1,Gutsche13a}
\ba
F_{1N}^\ast &=& 
\frac{1}{12 \sqrt{2}} 
(1 + g_V ) \delta_N 
\frac{Q^4}{m_\rho^4} G_2 + \nonumber \\
& & 
\frac{1}{24}(c_1 + c_2 g_V) \delta_N \frac{Q^2}{m_\rho^2}  G_2  + \nonumber \\  
& &
\frac{1}{60} \eta_N
\left( 
2 \sqrt{2}  \frac{Q^4}{m_\rho^4} 
 - c_3  \frac{Q^2}{m_\rho^2} + c_4 
\right)  \frac{Q^2}{m_\rho^2} 
G_3, 
\label{eqF1R}
%\ea  \ba
\\
F_{2N}^\ast &=& 
\frac{\sqrt{3}}{4}  \eta_N
\left( \frac{M_{N\!1} + M_N}{M_{N\!1}}
\right)^2 
\left( c_5 \frac{Q^2}{m_\rho^2} -4
\right) G_2,
\label{eqF2R}
\ea
where $\delta_N = \pm$ ($\delta_p=+1$, $\delta_n=-1$)
and  
the $c_i$ coefficients are 
$c_1=4 \sqrt{2} + 3 \sqrt{3}$, 
$c_2=4 \sqrt{2} - 3 \sqrt{3}$, 
$c_3= 9 (\sqrt{3} -\sqrt{2})$,
$c_4= 3 \sqrt{3}- 5 \sqrt{2}$
and $c_5=2 \sqrt{6}$.
As discussed next, 
the  functions $G_2$, $G_3$ can be written as
%The functions $G_2$, $G_3$ can be written as 
\ba
G_2 &=& 
\frac{1}{\left( 1 + \frac{Q^2}{m_\rho^2}\right)
\left( 1 + \frac{Q^2}{m_{\rho 1}^2}\right)
\left( 1 + \frac{Q^2}{M_{N}^2}\right)
\left( 1 + \frac{Q^2}{M_{N\!1}^2}\right)}, \nonumber \\
\label{eqG2} \\
G_3 &=& 
\frac{1}{\left( 1 + \frac{Q^2}{m_\rho^2}\right)
\left( 1 + \frac{Q^2}{m_{\rho 1}^2}\right)
\left( 1 + \frac{Q^2}{m_{\rho 2}^2}\right)
}
\nonumber \\
%\nonumber \ea
%\ba
& &
\times 
\frac{1}{
\left( 1 + \frac{Q^2}{M_N^2}\right)
\left( 1 + \frac{Q^2}{M_{N\!1}^2}\right)},
\label{eqG3}
\ea
where $M_N$, $M_{N\!1}$ are the mass of the 
nucleon and the nucleon first radial excitation (Roper),
and $m_\rho$, $m_{\rho n}$ ($n=1,2$)
are the $\rho$ mass and the mass of the 
first $\rho$ excitation.
All the masses are interpreted as empirical masses.

Equations (\ref{eqF1R}) and (\ref{eqF2R}) are derived in Ref.~\cite{paper1}
based on the expressions from Ref.~\cite{Gutsche13a}.
The difference between the present model and Ref.~\cite{paper1}
is that we represent here the functions 
$G_2$ and $G_3$ given by Eqs.~(\ref{eqG2}) and (\ref{eqG3})
in terms of the empirical masses of the hadrons.

The motivation for the present representation
comes from the expressions for the masses 
of the $\rho$-mesons and the nucleon radial excitations
derived  from light-front holography.
For the nucleon and $\rho$,  
the holographic expressions for the masses 
are $M_N= 2 \sqrt{2} \kappa$ and
$m_\rho= 2 \kappa$, respectively~\cite{Brodsky15,Abidin09}.
As for the $\rho$ excitations,  
we follow the parametrization $m_{\rho n}= 2 \kappa \sqrt{2 n + 1}$ 
for $n=1,2,...$,
derived in Refs.~\cite{Brodsky15,Teramond12a,Sufian17a},
where the twist-2 mass poles are shifted
to their physical values. 
A good description of the 
$\rho$ masses is obtained when $\kappa =0.385$ GeV (10\% error).

In the equations for $G_2$ and $G_3$ there is a term $\frac{Q^2}{M_{N\!1}^2}$,
where $M_{N\!1}$ is interpreted as the Roper mass ($M_R$).
In this case we replace $4 \kappa \simeq 1.52$ GeV by $M_{N\!1}$,
where  $M_{N\!1}$ is a rough estimation of the empirical mass of the
Roper $M_R \simeq 1.44$ GeV.
This replacement constitutes an exception 
to the identification of the 
hadron masses with the masses determined by light-front holography,
and it is justified by the result $M_R \simeq 4 \kappa$
(5\% deviation).
The corollary of the present parametrizations for $G_2$ and $G_3$
is the explicit dependence of the 
$\gamma^\ast N \to N(1440)$ transition form factors 
on the Roper empirical mass.

In Eqs.~(\ref{eqF1R}) and (\ref{eqF2R}) 
the only unknown parameters are 
$g_V$, $\eta_p$ and $\eta_n$.
In previous works~\cite{Gutsche12a,Brodsky15,Abidin09,Chakrabarti13a}
these parameters were determined by the 
nucleon data near the photon point.
However, since the coefficients $g_V$, $\eta_p$ and $\eta_n$
are related with the intrinsic properties of the valence quarks, 
we believe that these coefficients are more accurately determined 
by the large-$Q^2$ data, since in that case 
the effect of the meson cloud is significantly reduced.
In  Ref.~\cite{paper1} 
the nucleon elastic form factor data for $Q^2 > 1.5$ GeV$^2$
were studied using a light-front holographic model. 
It was concluded that the 
values $g_V=1.42\pm0.15$, $\eta_p = 0.39\pm 0.03$ and 
$\eta_n=- (0.34 \pm 0.02)$ provide a good description of the $Q^2 > 1.5$ GeV$^2$ 
data~\cite{ProtonData,RefsGEn,RefsGMn}.
In the present work, we consider the central values.
Since in the large-$Q^2$ region, the 
meson cloud effects are expected to be small, 
the values obtained for  $g_V$, $\eta_p$ and $\eta_n$
can be interpreted as bare couplings~\cite{paper1}.

The holographic estimates can also be compared 
with the transverse ($A_{1/2}$) and longitudinal 
($S_{1/2}$) amplitudes in the Roper rest frame.
Those helicity amplitudes  are related with the transition 
form factors through~\cite{NSTAR,Aznauryan07,Roper}
\ba
& &
A_{1/2}= {\cal R} (F_{1N}^\ast + F_{2N}^\ast), 
\label{eqA12}\\ 
& &
S_{1/2} = \frac{ {\cal R}}{\sqrt{2}} |{\bf q}|
\frac{M_R + M}{Q^2}( F_{1N}^\ast - \tau F_{2N}^\ast),
\label{eqS12}
\ea
where $\tau= \sfrac{Q^2}{(M_R+M)^2}$ and 
$|{\bf q}|$ is the photon three-momentum 
in the Roper rest frame. 
The factors ${\cal R}$ and $|{\bf q}|$ 
are respectively 
\ba
{\cal R}= \sqrt{\frac{\pi \alpha Q_-^2}{M_R M K}},
\hspace{.8cm}
|{\bf q}| = \frac{\sqrt{Q_+^2 Q_-^2}}{2M_R},
\ea
where $Q_\pm^2= (M_R \pm M)^2 + Q^2$, 
$K = \frac{M_R^2-M^2}{2M_R}$ and 
$\alpha \simeq 1/137$ is the fine structure constant.

\begin{figure}[t]
\vspace{.7cm}
%\mbox{
\includegraphics[width=3.1in]{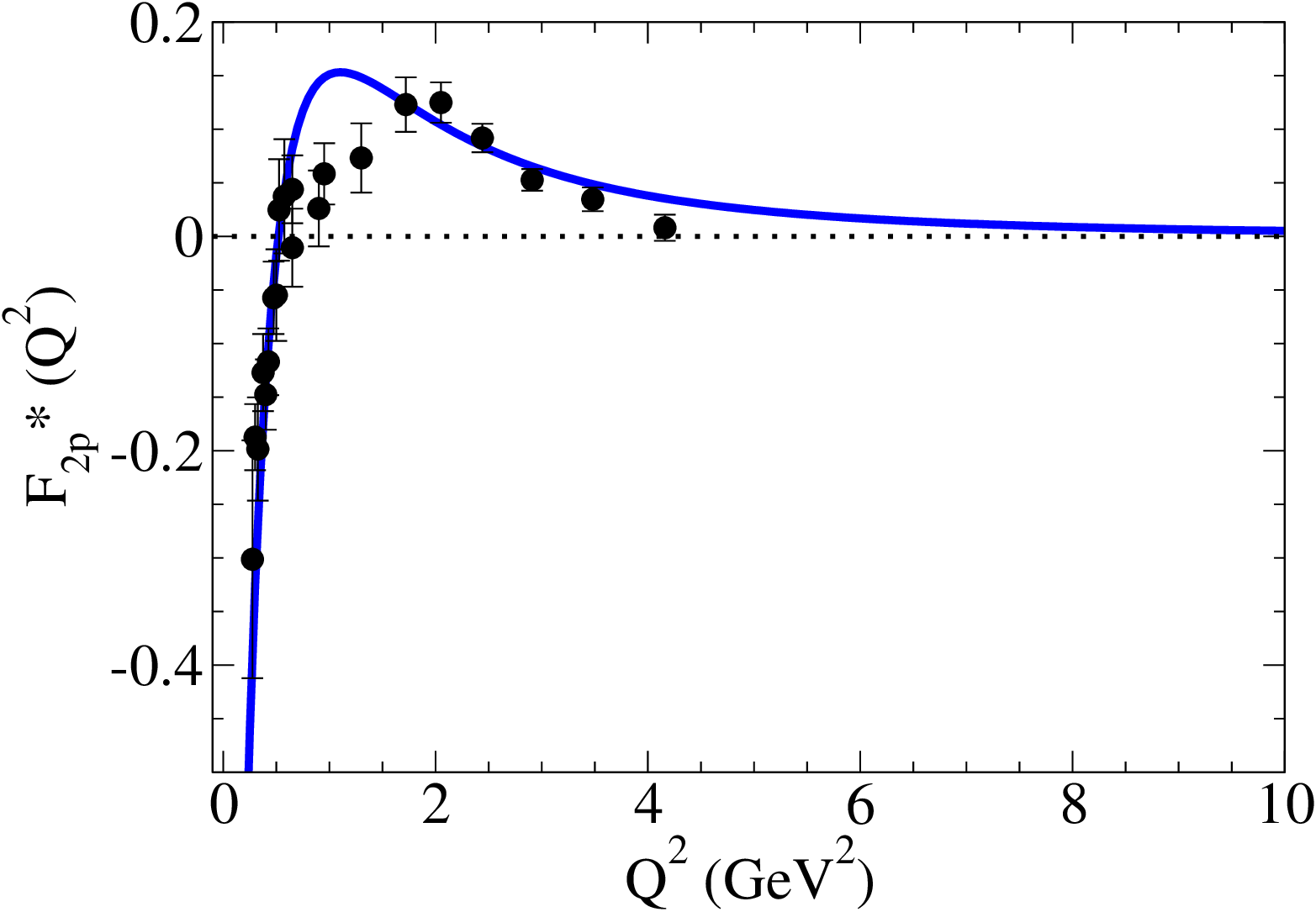}%}
%\mbox{
\vspace{.9cm}
\includegraphics[width=3.1in]{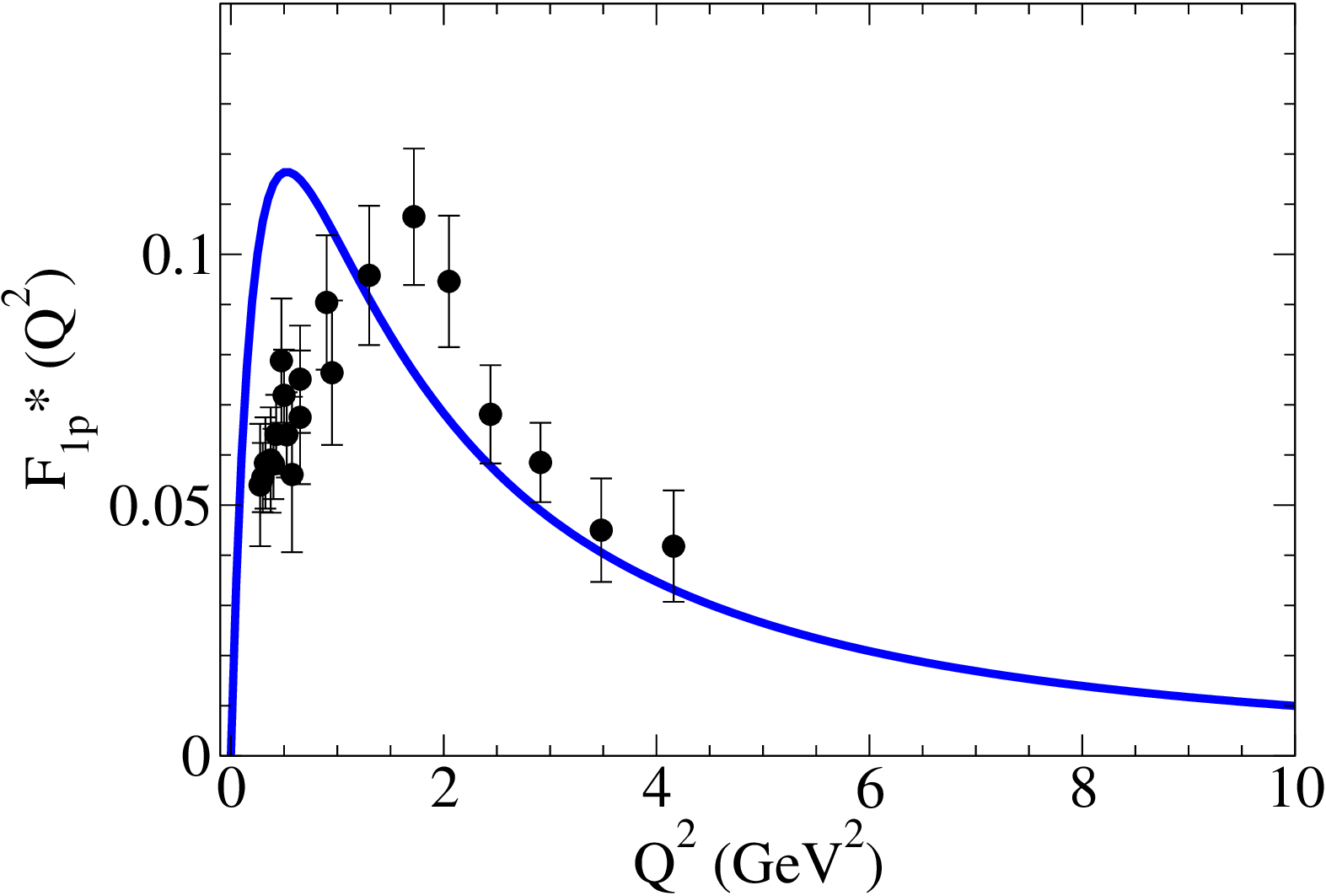}%}
\caption{\footnotesize{
$\gamma^\ast N \to N(1440)$ transition 
form factors $F_{1p}^\ast$ and  $F_{2p}^\ast$.
Data from Refs.~\cite{CLAS1,CLAS2}.
}}
%\vspace{-1cm}
\label{figRoper}
\end{figure}

\begin{figure}[t]
\vspace{.5cm}
%\mbox{
\includegraphics[width=3.1in]{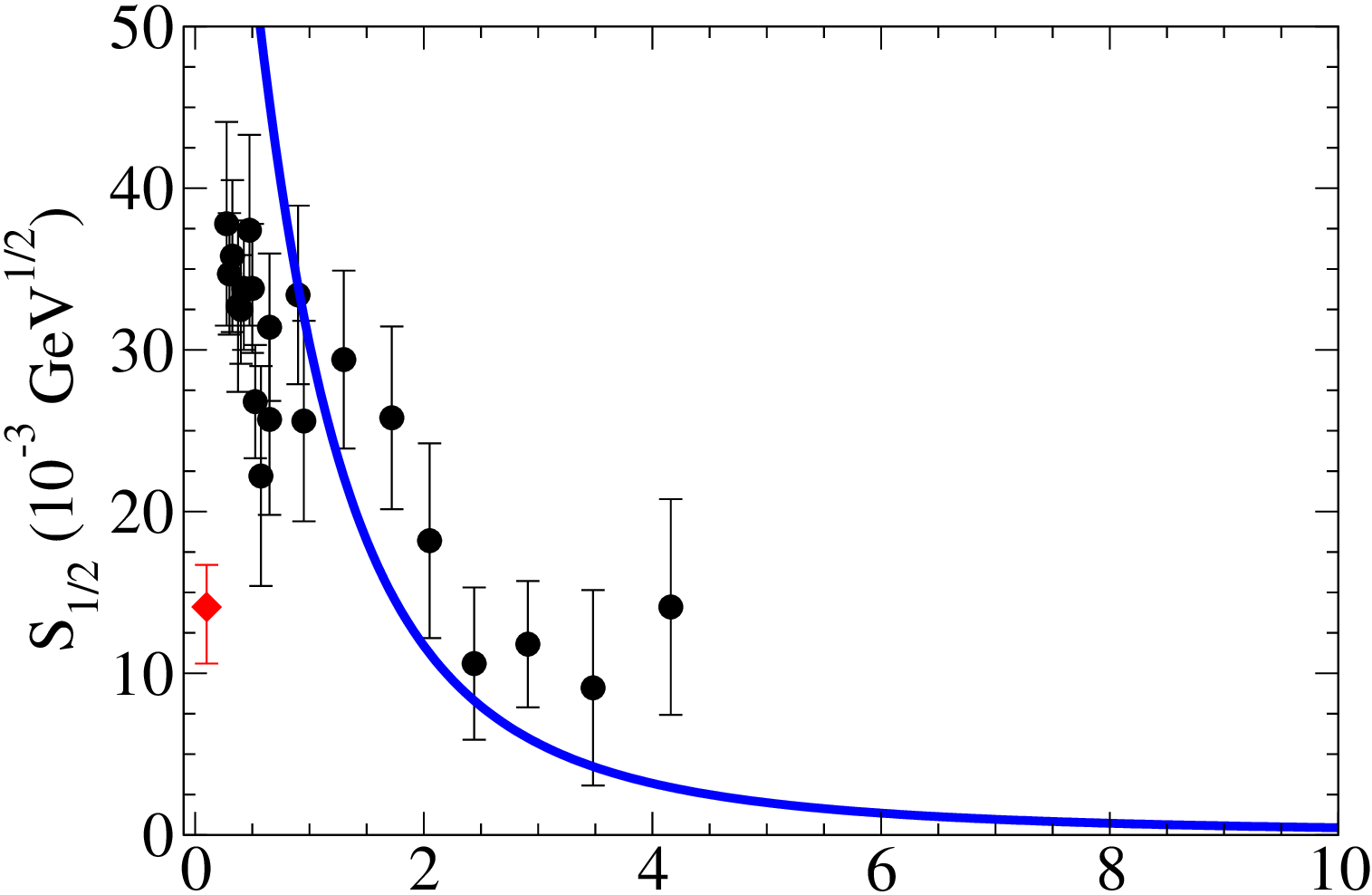}%}
%\mbox{
\vspace{.8cm}
\includegraphics[width=3.1in]{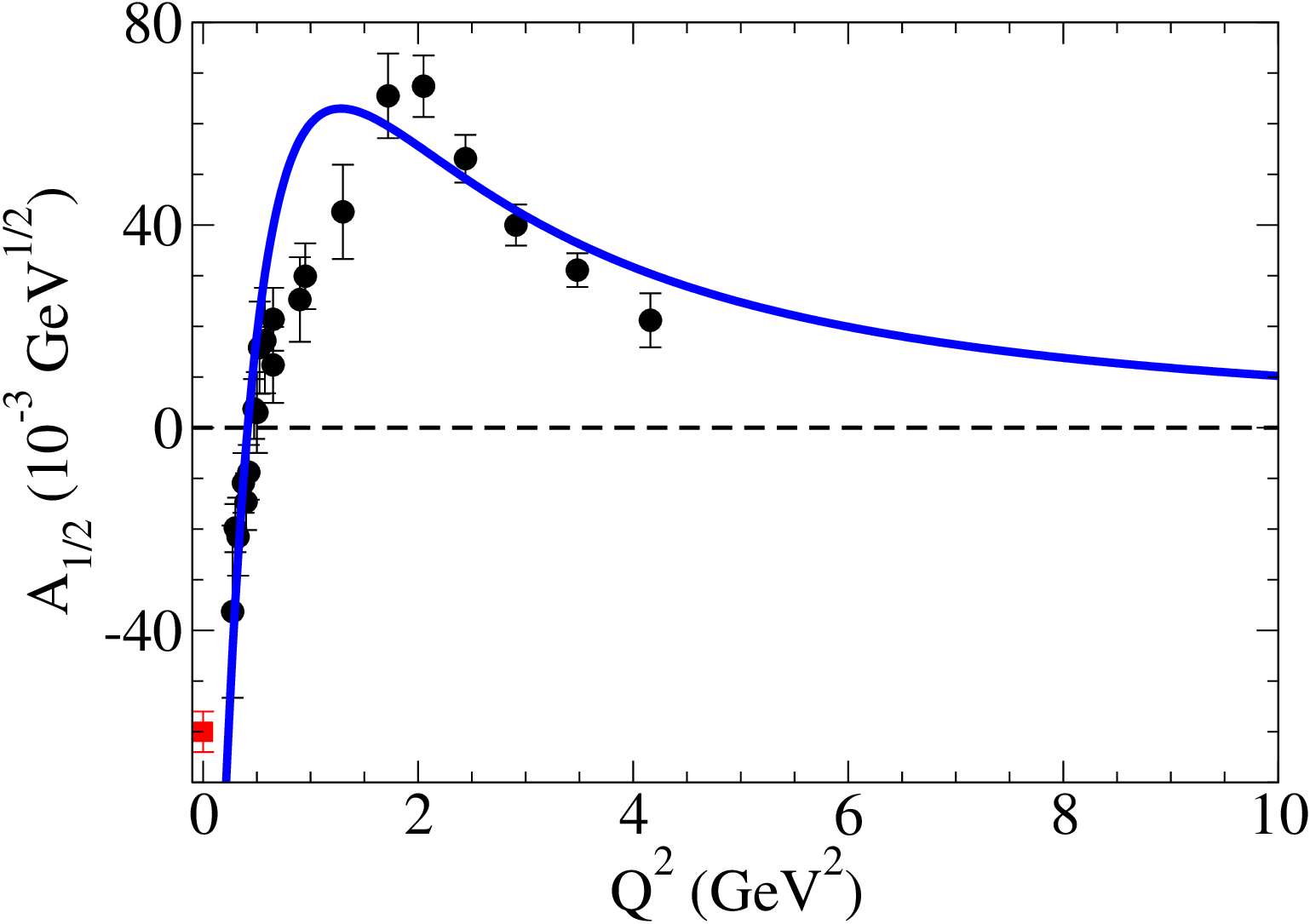}%}
\caption{\footnotesize{
$\gamma^\ast N \to N(1440)$ helicity amplitudes.
Data from CLAS~\cite{CLAS1,CLAS2} (solid circles), 
Particle Data Group~\cite{PDG} for $A_{1/2}$ (diamond)
and A1 collaboration~\cite{Stajner17a} for $S_{1/2}$ (square).
}}
%\vspace{-1cm}
\label{figRoperAmp}
\end{figure}

The results for the nucleon to Roper transition 
for the proton target ($N = p$) are presented in 
Figs.~\ref{figRoper} and \ref{figRoperAmp}, 
for the form factors and helicity amplitudes, respectively.
The model estimates are compared 
with the data from CLAS/Jefferson Lab~\cite{CLAS1,CLAS2}.
For the amplitudes we include also the 
result for $A_{1/2}$ at the photon point from 
Particle Data Group~\cite{PDG}
and the recent result for $S_{1/2}$ for $Q^2 \simeq 0.1$ GeV$^2$ 
from the A1 collaboration~\cite{Stajner17a}.
We do not discuss the results for the
neutron target ($N = n$), since one has data only for $Q^2=0$.


The results presented in Fig.~\ref{figRoper}
for the transition form factors
are very interesting because they show 
that the experimental data can be described by 
simple parametrizations 
based on the empirical masses of the hadrons.
The agreement between the parametrization 
and the data is very good particularly 
for $Q^2 > 2$ GeV$^2$, 
where the valence quark degrees of freedom are dominant.

For low $Q^2$, we should not expect  
the parametrizations to be so accurate, because 
as mentioned, Eqs.~(\ref{eqF1R}) and (\ref{eqF2R}) 
do not include the contributions from the meson cloud, 
which may be in general significant for small $Q^2$,
and also because the bare parameters $g_V$, $\eta_p$ and $\eta_p$
are estimated by the large-$Q^2$ nucleon data~\cite{paper1}.
In the upper panel of Fig.~\ref{figRoper}, we present our 
estimates to the Roper Pauli form factor $F_{2p}^\ast$.
As far as we know, this is the first time that 
a simple analytic expression is presented for $F_{2p}^\ast$.
Although the calculations are expected to be valid only
for large $Q^2$, it is nevertheless interesting to note
the remarkable description of the function $F_{2p}^\ast$
in the region 0.2--3.5 GeV$^2$.

From the results for the helicity amplitudes, presented in Fig.~\ref{figRoperAmp}, 
we can confirm that those parametrizations are also accurate for $Q^2 > 1.5$ GeV$^2$.
This conclusion could have been anticipated from the analysis of the form factors.
For low $Q^2$ we can observe some deviations from the data,
particularly for $S_{1/2}$.
This result is mainly the consequence of the model overestimation 
of the function  $F_{1p}^\ast$ for $Q^2 < 1$ GeV$^2$, 
since the effect of $F_{2p}^\ast$ is reduced by the factor $\tau$,
according with Eq.~(\ref{eqS12}).
As for the amplitude $A_{1/2}$, the deviation 
from the data happens only near $Q^2=0$,
as a consequence of the results for the function $F_{2p}^\ast$,
since $F_{1p}^\ast(0)=0$.
In both cases the deviation from the data  at low $Q^2$
can be the result of no inclusion of 
meson cloud contributions,
or a limitation of the holographic formalism.
Recall that the holographic estimates 
are calculated in the limit of zero quark masses.

In the graph for $S_{1/2}$ 
it is possible to note the small value 
obtained at $Q^2 \simeq 0.1$ GeV$^2$ by the A1 collaboration~\cite{Stajner17a}.
This result may be an indication that the amplitude $S_{1/2}$ vanishes 
at the pseudothreshold, when $Q^2= -(M_R-M)^2 \simeq -0.25$ GeV$^2$,
according to the long-wavelength 
constraint~\cite{Stajner17a,AmaldiBook,Tiator1,Tiator2,Siegert1,Siegert2}.

Once confirmed the accuracy 
of the parametrizations from Eqs.~(\ref{eqF1R}) and (\ref{eqF2R}) 
for the $\gamma^\ast N \to N(1440)$ 
Dirac and Pauli transition form factors, 
it is worth checking if  for the case of the nucleon 
one can also obtain simple analytic expressions 
based on the masses of hadrons, 
and if those parametrizations 
describe well  the nucleon form factor data.

\begin{figure}[t]
\vspace{.5cm}
\centerline{
\mbox{
\includegraphics[width=1.65in]{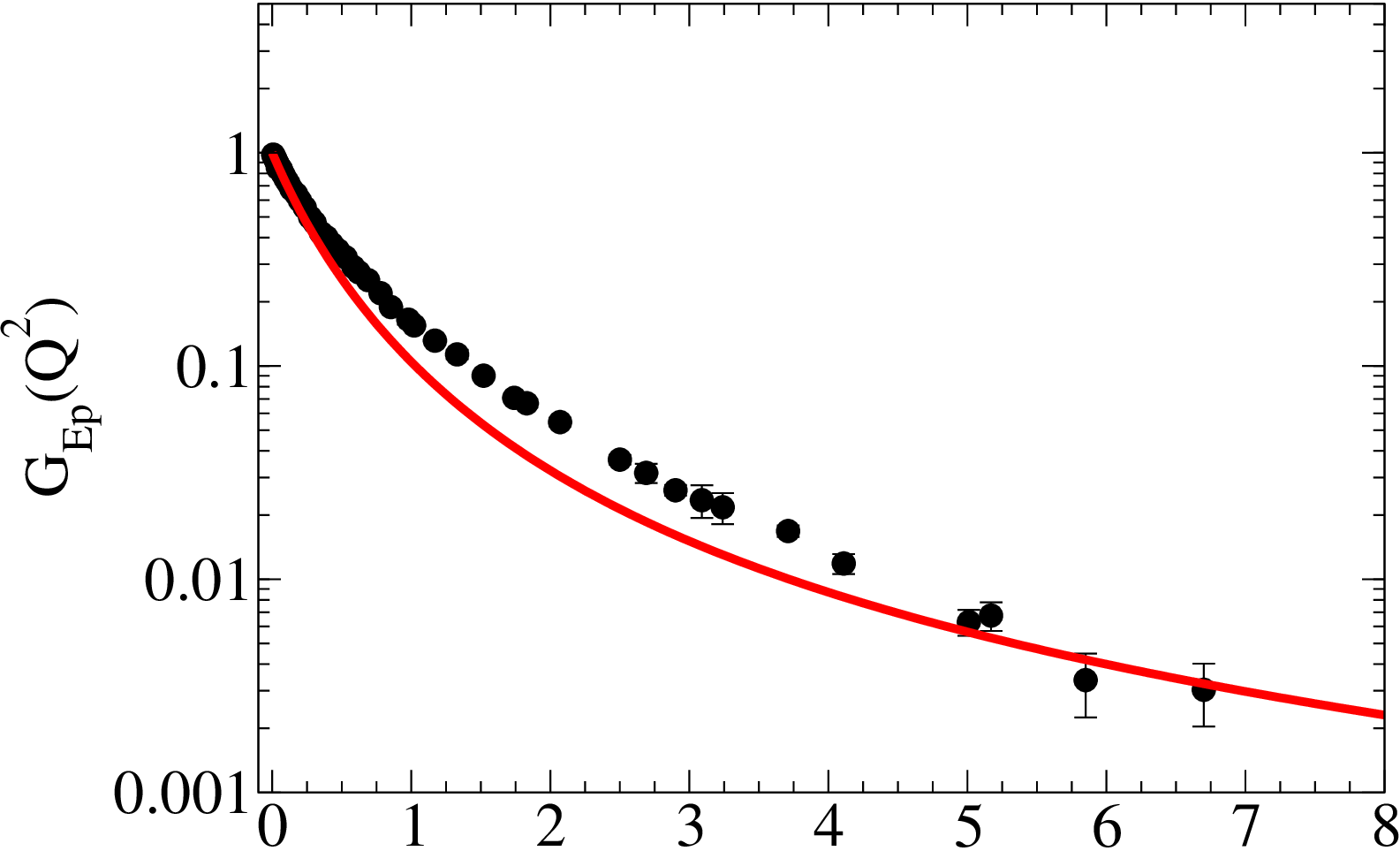}  \hspace{.1cm}
\includegraphics[width=1.65in]{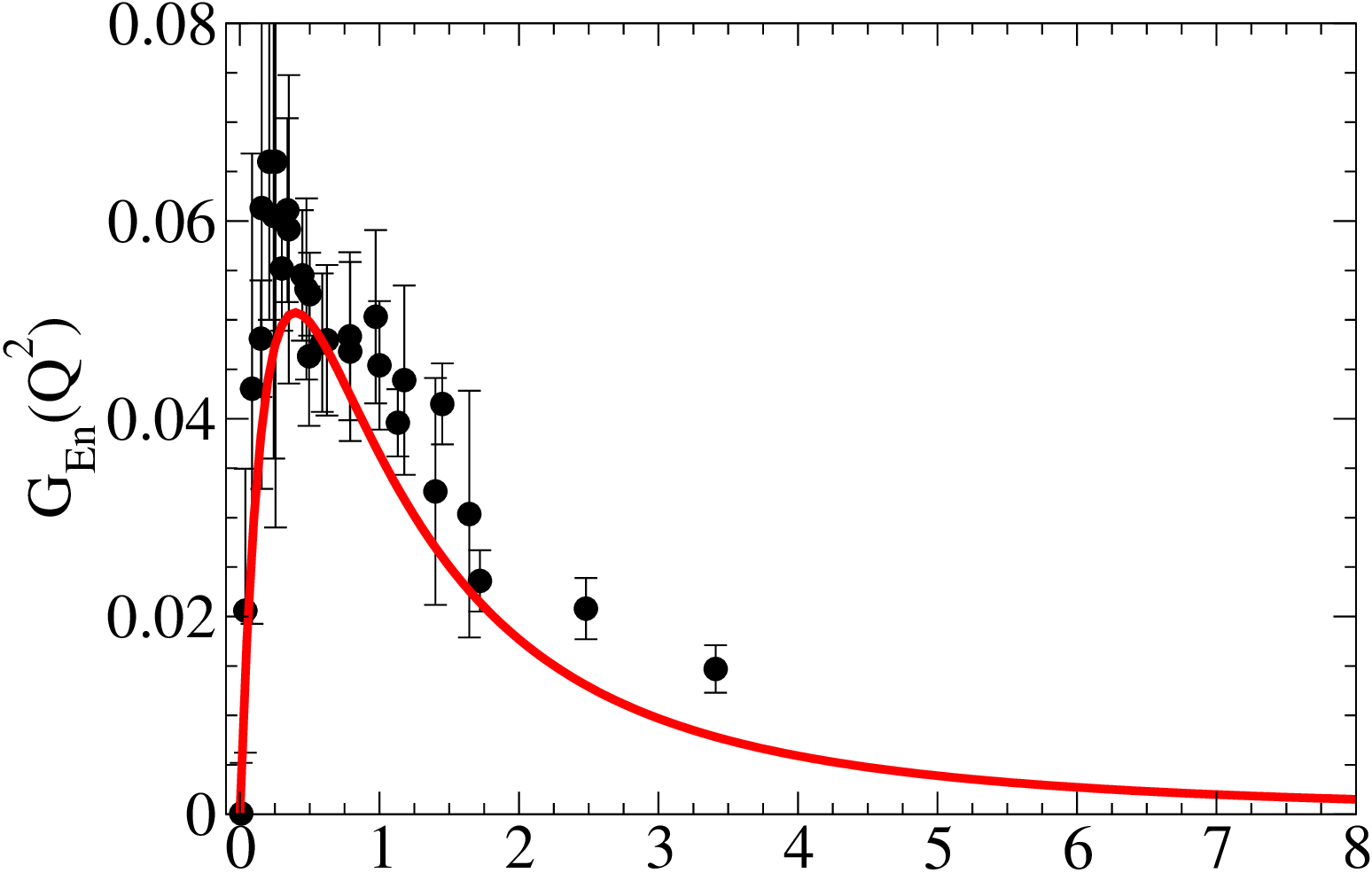}
}}
\vspace{.6cm}
\centerline{
\mbox{
\includegraphics[width=1.65in]{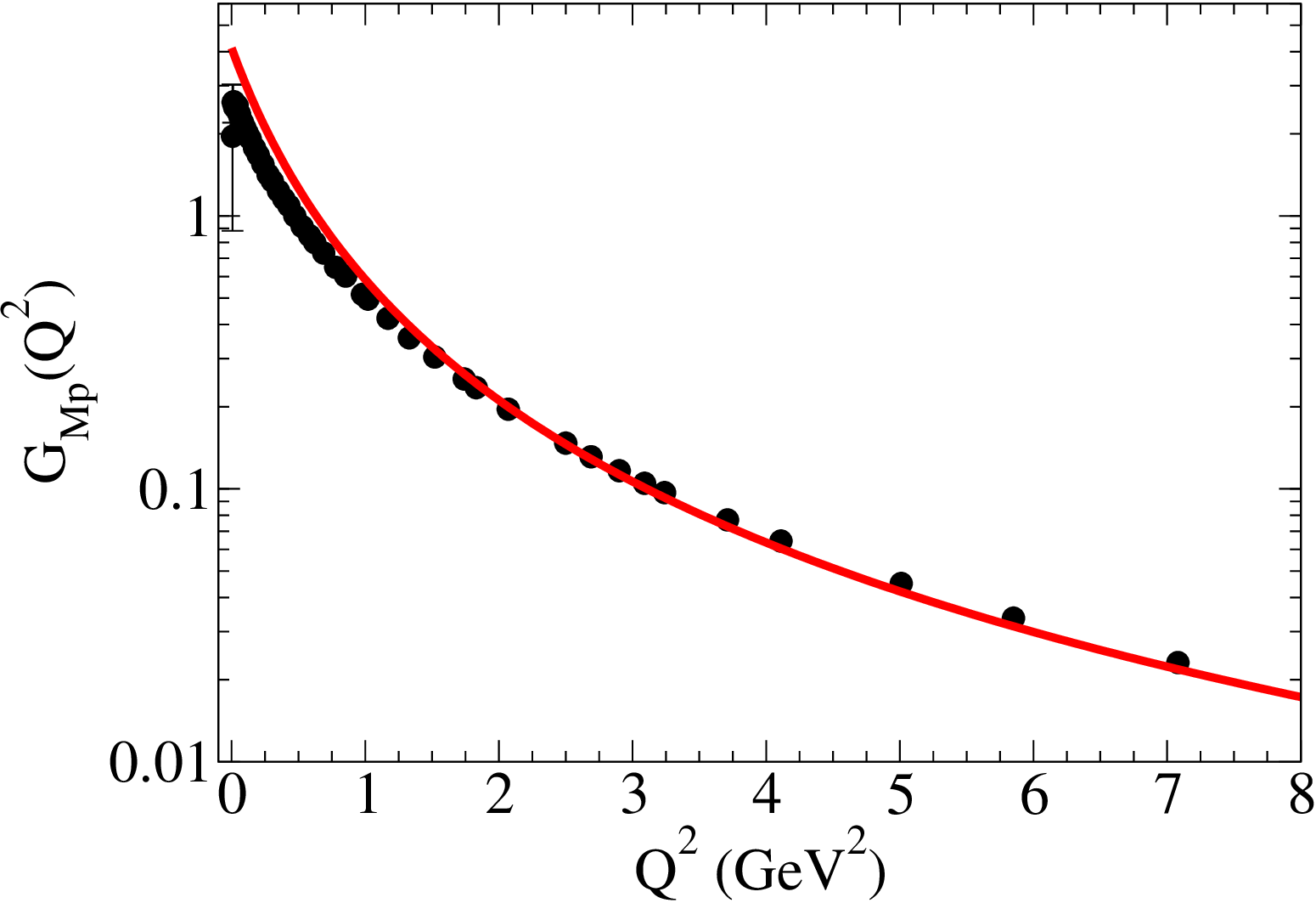}  \hspace{.1cm}
\includegraphics[width=1.65in]{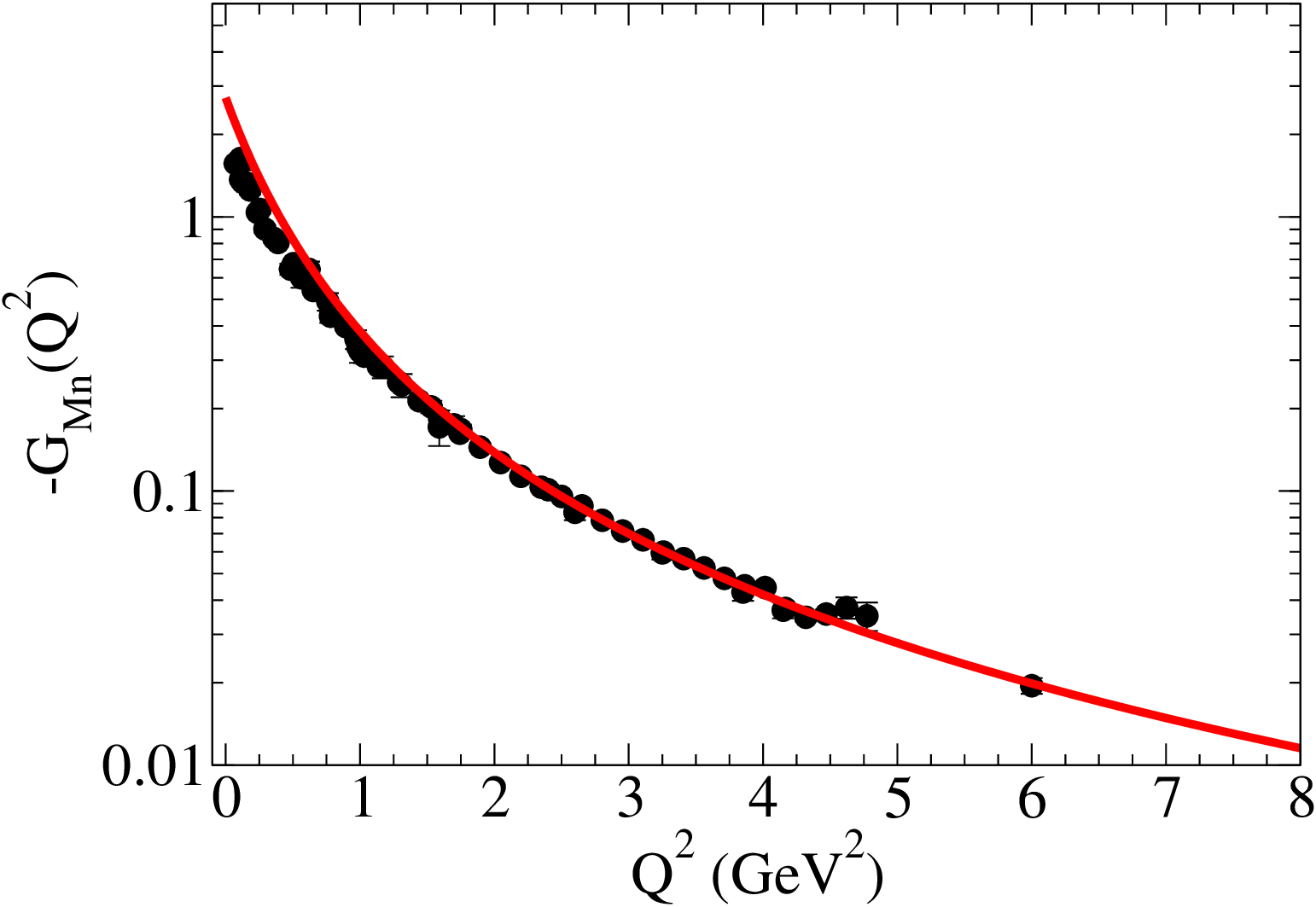}
}}
\caption{\footnotesize{
Proton and neutron electric and magnetic form factors.
Data from Refs.~\cite{ProtonData,RefsGEn,RefsGMn}.
}}
%\vspace{-1cm}
\label{figGEGMX}
\end{figure}

\subsection*{Nucleon elastic form factors}

The nucleon electromagnetic form factors 
can be parametrized as 
\ba
F_{1N}&=& 
e_N G_1 + \frac{1}{6} (e_N + \delta_N g_V) \frac{Q^2}{m_\rho^2} 
 G_1 \nonumber \\
& &  -   \frac{1}{6} \eta_N \frac{Q^2}{m_\rho^2} 
\left(
\frac{1}{2} -   \frac{Q^2}{m_\rho^2} 
\right) G_2, \label{eqF1N} \\
F_{2N} &=& 8 \eta_N G_1,
\label{eqF2N} 
\ea
where $e_N$ is the nucleon charge and 
\ba
G_1 &=& 
\frac{1}{\left( 1 + \frac{Q^2}{m_\rho^2}\right)
\left( 1 + \frac{Q^2}{m_{\rho 1}^2}\right)
\left( 1 + \frac{Q^2}{M_{N}^2}\right)}. 
\label{eqG1} 
\ea

Equations (\ref{eqF1N})-(\ref{eqF2N}) are obtained from 
Ref.~\cite{paper1} replacing the 
holographic nucleon mass ($2\sqrt{2} \kappa$) by the  
the nucleon physical mass,  
combined with the re-interpretation of the
poles of the function $G_1$ as the empirical hadron 
masses ($m_\rho$, $m_{\rho 1}$ and $M_N$), 
as discussed for the case of the Roper.

The results associated with the previous 
analytic expression 
for the electric $G_{E\!N} = F_{1\! N} - \sfrac{Q^2}{4 M_N^2} F_{2\!N}$
and magnetic $G_{M\!N} = F_{1\! N} + F_{2\!N}$ form factors
are presented in Fig.~\ref{figGEGMX} up to 8 GeV$^2$.
We use a logarithm scale for 
$G_{Ep}$, $G_{Mp}$ and $G_{Mn}$ in order 
to better observe the falloff of those form factors 
for larger values of $Q^2$.

\begin{figure}[t]
\vspace{.5cm}
%\mbox{
\includegraphics[width=3.0in]{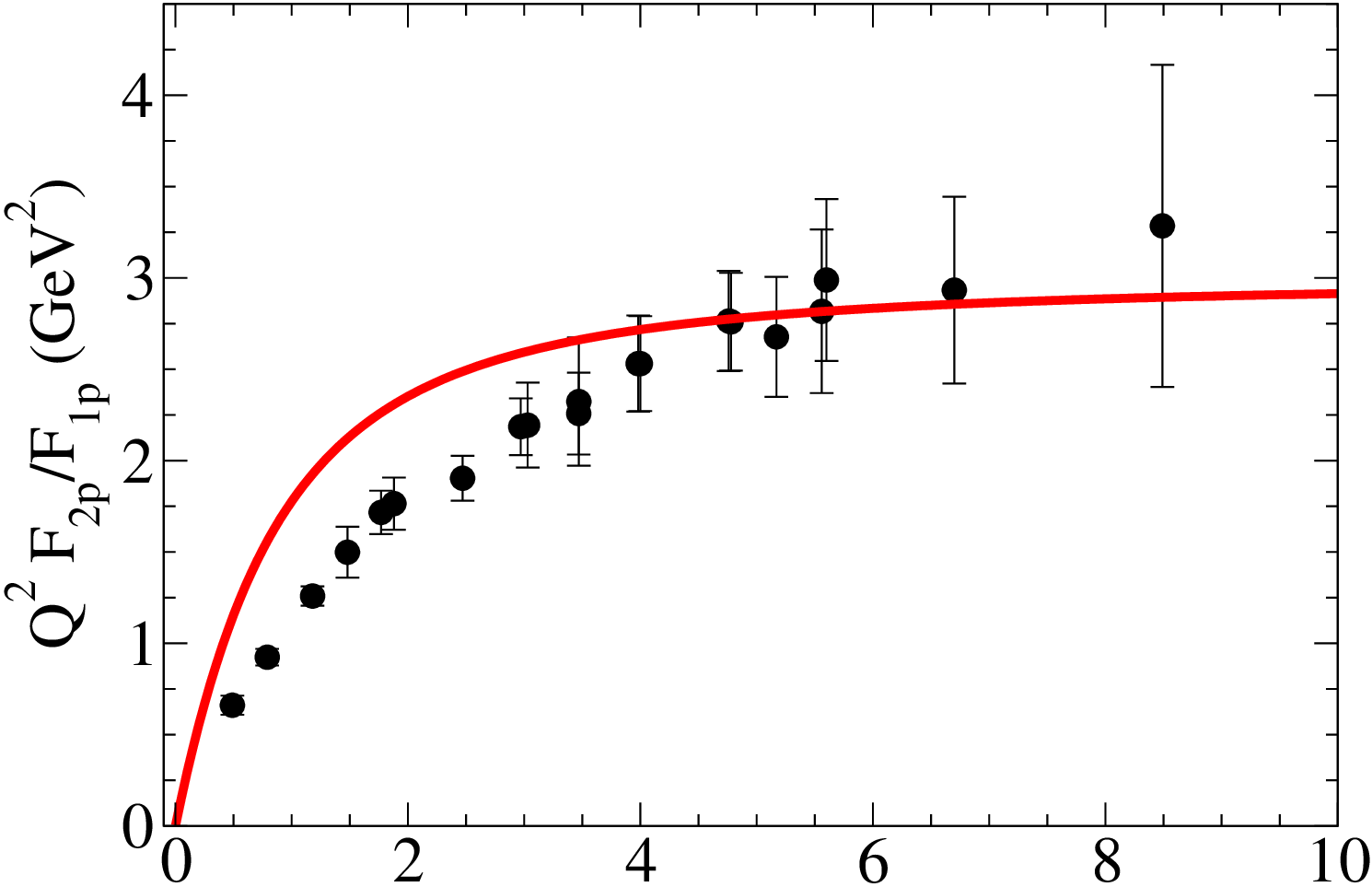}%}
%\mbox{
\vspace{.9cm}
\includegraphics[width=3.0in]{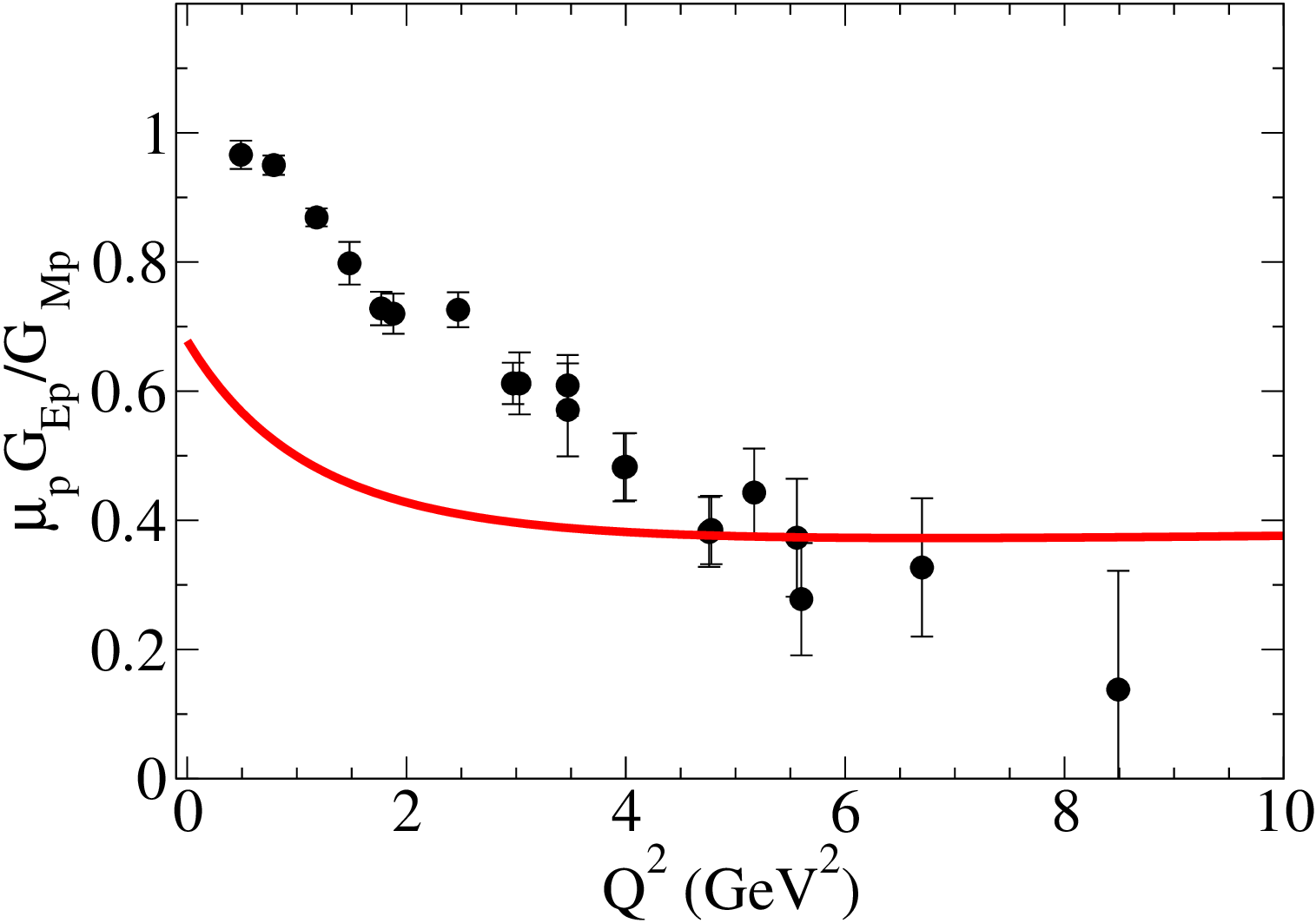}%}
\caption{\footnotesize{
Ratios between proton electromagnetic form factors.
Data from Ref.~\cite{ProtonData}.
}}
%\vspace{-1cm}
\label{figProton}
\end{figure}

In Fig.~\ref{figGEGMX} one can see that 
Eqs.~(\ref{eqF1N}) and (\ref{eqF2N}) 
provide accurate representations 
of the nucleon form factor data  for $Q^2 > 1.5$ GeV$^2$,
except for the proton electric form factor ($G_{Ep}$),
where the agreement with the data happens only for $Q^2 > 4$ GeV$^2$.
A better description of $G_{Ep}$ can be obtained 
if we take into account  in $F_{2p}$
the difference between the nucleon physical 
and the holographic estimate (15\% correction)~\cite{paper1}.
Considering all the form factors,
the failure for $Q^2 < 1.5$ GeV$^2$ may 
have been anticipated, 
since the bare parameters ($g_V$, $\eta_p$ and $\eta_n$) 
are determined from fits to the large-$Q^2$ data.

The results for $G_{Ep}$ suggest that the meson cloud 
effect may play a more important role in this specific form factor.
$G_{Ep}$ is particularly sensible 
to the meson cloud effect, because in the case of the proton, 
the effect modifies the charge of the three valence-quark component,
which is reduced by a factor $Z_N < 1$, 
due to the normalization of 
the nucleon dressed wave function~\cite{NucleonMC1,NucleonMC2,AxialFF}.
In that case, the meson cloud component 
contributes with the charge $(1-Z_N)e$, where $e$ is 
the elementary charge.

In principle, the description of the low $Q^2$ region 
can be improved with the inclusion of high order Fock 
states, in particular with the term $\tau=5$
as in Ref.~\cite{Sufian17a}. 
In the present work, however, our focus 
is in the derivation of simple analytic expressions 
for the  $\gamma^\ast N \to N(1440)$ transition form factors,
valid for large $Q^2$, which are expected to 
be dominated by the valence quark contributions.

Concerning the proton form factors, 
one can compare our results directly with 
the ratio $\mu_p G_{Ep}/G_{Mp}$, which can be measured 
by polarization transfer experiments at Jefferson Lab~\cite{NSTAR,ProtonData}.
The comparison with the data analysis from Refs.~\cite{ProtonData}
is presented in the lower panel of Fig.~\ref{figProton}.
From that data one can also estimate the ratio $F_{2p}/F_{1p}$ 
scaled by the factor $Q^2$. 
The results are presented in the upper panel of Fig.~\ref{figProton}.

In both cases one can observe a deviation from 
the data below 3 GeV$^2$ and a good agreement for larger values of $Q^2$.
The discrepancy for $\mu_p G_{Ep}/G_{Mp}$ is a consequence 
of the discrepancies noted already for $G_{Ep}$ and $G_{Mp}$
(underestimation for $G_{Ep}$ and overestimation for $G_{Mp}$).
The highest datapoint ($Q^2 \simeq 8.5$ GeV$^2$) 
differs from the model estimate only by 1.3 standard deviations.
For $Q^2 > 4$ GeV$^2$, one can observe an almost perfect 
scaling between $Q^2 F_{2p}$ and $F_{1p}$.

From the results, one concludes that   
the nucleon elastic form factors at large $Q^2$ 
can also be described by simple analytic expressions  
dependent on the bare couplings, 
the nucleon mass and the $\rho$ meson masses.
The parametrization for the neutron electric 
form factor, in particular, 
may be tested in the near future, 
once the results from the JLab-12 GeV upgrade become available 
for $Q^2 > 4$ GeV$^2$~\cite{NSTAR}.

\section{Summary and conclusions}

In the present work we use previous results from  
light-front holography to derive analytic 
parametrizations of the nucleon and 
$\gamma^\ast N \to N(1440)$ form factors   
in leading twist approximation.
In the leading twist approximation the nucleon 
and the nucleon excitations are described 
as three valence-quark systems.
The falloff of the Dirac and Pauli form factors 
are consistent with the behavior expected 
from perturbative QCD~\cite{paper1,Carlson}.   

The parametrizations presented here are based on 
two main ingredients: 
(i) the couplings associated with 
the electromagnetic interaction to the quarks 
are interpreted as bare couplings 
(no meson cloud contamination);
(ii) the analytic structure of the 
Dirac and Pauli form factors may be represented 
in terms of the masses of the vector mesons 
($\rho$ mesons) and the masses of the nucleon 
and the nucleon first radial excitation (Roper).

In particular the parametrizations associated with the Roper form factors,
based on Eqs.~(\ref{eqF1R})-(\ref{eqF2R}),
give a very good description  of the data for $Q^2 > 2$ GeV$^2$, 
corroborating the idea that, 
in fact, the resonance $N(1440)$ is the first radial 
excitation of the nucleon, as suggested by  
several authors~\cite{Aznauryan07,Roper,Segovia15}.
At low $Q^2$, however, the effect of the meson cloud effects cannot be ignored.
Meson cloud effects are also fundamental to 
explain the mass of the Roper and 
its decay widths~\cite{Segovia15,Suzuki10,Gegelia16}.

A simple analytic expression
was previously proposed for the Roper Dirac 
form factor~\cite{Brodsky15,Teramond11a},
in close agreement with the large-$Q^2$ data.
However, to the best of our knowledge, 
this is the first time that
a simple analytic expression is presented for the Roper Pauli form factor.
The deviations from the data at low $Q^2$ 
can be interpreted as the result 
of the meson cloud effects, which are omitted  
in leading twist approximation.
Surprisingly, our estimate of  
the Roper Pauli form factor is also very 
accurate for small $Q^2$, more specifically 
in the region $Q^2=0.2$--1 GeV$^2$.

Light-front holography may in the future 
be used for the study of other $\gamma^\ast N \to N^\ast$ transitions, 
where $N^\ast$ is a generic $J^P$ resonance (spin $J$ and parity $P$)
in leading twist approximation.
The parameters associated with the electromagnetic interaction
are already fixed by the study of the nucleon form factors 
and were successfully tested for the case of the Roper.
One can then expect that analytic expressions 
for the transition form factors may also be derived for  
other $\gamma^\ast N \to N^\ast$ transitions.
It is also likely that new analytic expressions 
for the transition form factors based on the empirical hadron masses 
may be derived.

%\newpage

To summarize, the formalism presented 
in this work for the Roper can
in the near future be extended to higher mass nucleon excitations.
The use of the light-front holography in leading
twist approximation provides a natural method to estimate the valence
quark contributions for the transition form factors.
The effect of the meson cloud can then be estimated from
the comparison with the experimental data.
In addition the light-front holographic parametrization may be tested 
in the near future by the results from the
JLab 12-GeV upgrade~\cite{NSTAR} for $Q^2 > 6$ GeV$^2$,
a region where we expect the estimates to be very accurate.

\begin{acknowledgments}
%\vspace{.2cm}
%{\bf Acknowledgments:}
The author thanks Dmitry Melnikov for useful discussions
and Pulak Giri for helpful suggestions.
This work was supported by the
Universidade Federal do
Rio Grande do Norte/Minist\'erio da Educa\c{c}\~ao (UFRN/MEC). 
\end{acknowledgments}

\end{document}